\documentclass[journal]{IEEEtran}
\usepackage{graphicx}

\begin{document}
\title{Performance of Several Solid State Photomultipliers with CLYC Scintillator}

\author{Katherine~E.~Mesick, Laura~C.~Stonehill, Jonathan~T.~Morrell, Daniel~D.S.~Coupland%
\thanks{Manuscript received December 3, 2015. This work has been supported by the US Department of Energy National Nuclear Security Administration Office of Defense Nuclear Nonproliferation Research and Development.}%
\thanks{K.E. Mesick, L.C. Stonehill, and D.D.S Coupland are with the ISR-1 Space Science and Applications Group at Los Alamos National Laboratory, Los Alamos, NM, 87545 USA (corresponding author: kmesick@lanl.gov).}%
\thanks{J.T. Morrell is with Massachusetts Institute of Technology.}%
}

\maketitle
\pagestyle{empty}
\thispagestyle{empty}

\begin{abstract}
Cs$\mathbf{_2}$LiYCl$\mathbf{_6}$:Ce$\mathbf{^{3+}}$ (CLYC) is an inorganic scintillator that has recently garnered attention for its ability to detect and discriminate between gammas and thermal neutrons.  We investigate several important performance parameters of three different solid state photomultipliers (SSPMs) when reading out CLYC crystals: linearity, energy resolution, and pulse shape and discrimination ability.  These performance parameters are assessed at a variety of temperatures between $\mathbf{-20^{\circ}}$C and $\mathbf{+50^{\circ}}$C.
\end{abstract}


\section{Introduction}
\IEEEPARstart{T}{he} recent development of Cs$_2$LiYCl$_6$:Ce$^{3+}$ (CLYC) \cite{clyc1} has gained much interest for its use as a dual gamma and neutron detector.  CLYC has a density of 3.3~g/cm$^3$, emits at a peak wavelength of about 373~nm, and provides excellent linearity and gamma energy resolution (full-width half maximum, FWHM) as good as 4\% at 662~keV and thermal neutron sensitivity through the $^6$Li(n,$\alpha$)T capture reaction \cite{clyc2}.  Different characteristic time response for incident gammas and thermal neutrons results in the ability to perform pulse-shape discrimination (PSD).

Scintillators are typically coupled to traditional photomultiplier tubes (PMTs) for readout.  However, solid state photomultipliers (SSPMs) offer potential advantages in certain applications due to having a smaller size, lower voltage requirement, added robustness, and being impervious to magnetic fields.  The variance of waveforms and discrimination ability with temperature for CLYC coupled to a PMT has previously been studied \cite{buddentemp}.  Here we study the linearity, gamma energy resolution, and pulse shape and discrimination ability over a range of temperatures from $-20^{\circ}$C to $+50^{\circ}$C for CLYC coupled to three different SSPMs.

\section{Experiment Method}

Three SSPMs were coupled to 1~cm$^3$ CLYC scintillators: a SensL MicroFC-SMTPA-60035 silicon photomultiplier (SiPM), a Hamamatsu Multi-Pixel Photon Counter (MPPC) S13083-050CS, and a KETEK PM6660TP-BS0 SiPM.  Each of these SSPMs has an active area of $6\times6$~mm$^2$ and was coupled directly to the CLYC crystal and hermetically sealed by Radiation Monitoring Devices Inc. (RMD).  The measured gamma resolution at 662~keV before delivery was 6.6\%, 8.3\%, and 6.2\% for the crystals coupled to the SensL, Hamamatsu, and KETEK, respectively.  All three crystals exhibited a resolution of $\sim$4\% when coupled to a Hamamatsu R6233 PMT prior to packaging.

For each SSPM, data were collected in three different ways: with a standard shaping amplifier and multi-channel analyzer (MCA) setup, with custom electronics capable of three different gated charge integrals over different portions of the scintillation pulse, and with a fast waveform digitizer.

The linearity and energy resolution for each system was characterized with three radioactive gamma sources read out with an Amptek 8000D MCA: $^{22}$Na (511~keV, 1275~keV), $^{137}$Cs (662~keV), and $^{57}$Co (122~keV).  The photopeaks were fit with a Gaussian plus linear background to extract the centroid and widths for each peak.

\begin{figure}[b]
\centering
\includegraphics[width=0.42\textwidth]{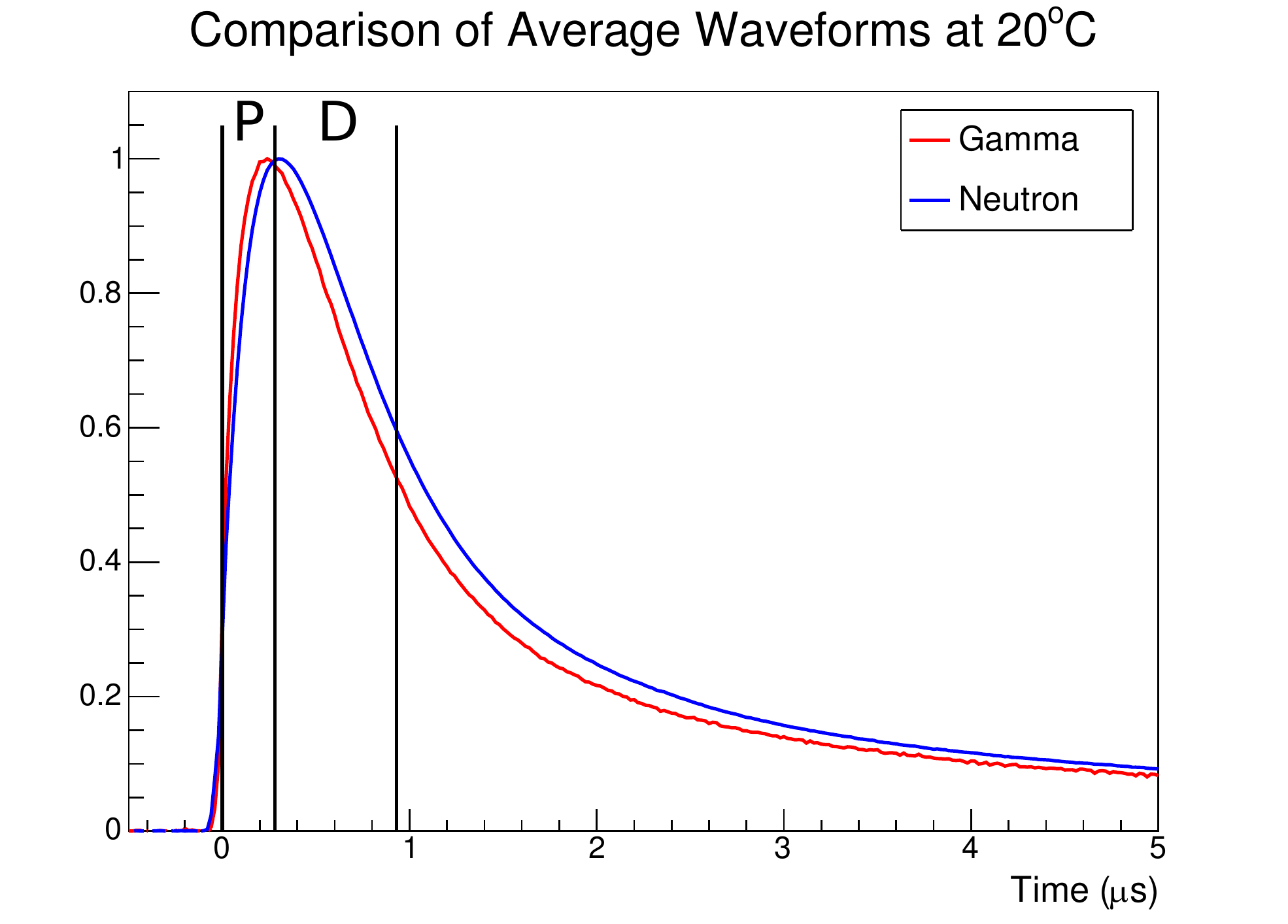}
\caption{Example of average waveforms at 20$^{\circ}$C with the determined prompt and delayed integration windows.}
\label{fig:psd}
\end{figure}

A moderated AmB gamma and neutron source was used to study pulse shape and discrimination ability.  PSD is performed by taking the ratio of delayed (D) and prompt (P) integration windows, seen for example in Fig.~\ref{fig:psd}, of the waveforms:
\begin{equation}
\textrm{PSD Ratio} = \frac{D}{D+P}~.
\end{equation}
\hspace{-2.5pt}In CLYC the neutron pulse has a slower recovery time, leading to a larger PSD ratio value than the gammas.  The delayed and prompt integration windows were optimized for each SSPM to maximize the figure of merit (FOM), defined as
\begin{equation}
\textrm{FOM} = \frac{\mu_n - \mu_g}{\Gamma_n + \Gamma_g}~,
\end{equation}
where $\mu_{g(n)}$ is the centroid and $\Gamma_{g(n)}$ the FWHM of the gamma (neutron) peak in the PSD ratio.  These optimized integration windows were programmed into the custom ASIC system for studying the FOM over the temperature range.  Pulse shapes were also recorded with an Agilent Acqiris DC282 waveform digitizer sampling at 50 MS/s.  Averaged waveforms were obtained by defining gamma and neutron cuts based on the PSD ratio and full-pulse integrated energy within a 10~$\mu$s window.

A thermal chamber was used to thermal cycle each CLYC crystal and attached SSPM from $-20^{\circ}$C to $+50^{\circ}$C in 10$^{\circ}$C increments.  The cycling procedure followed was derived from ANSI standard N42.34 for handheld radiation detectors \cite{ansin42}.  Only the crystal and SSPMs were in the thermal chamber; the radioactive sources and electronics were outside the thermal chamber during data collection.  The PSD integration windows, optimized at room temperature, were kept constant over the range of temperatures.  The bias voltage, set to 56~V for the Hamamatsu MPPC and 27~V for the SensL and KETEK SiPMs, was also kept fixed during temperature variation.

\section{Results}

\subsection{Linearity}\label{sec:lin}

\begin{figure}[h]
\centering
\includegraphics[width=0.42\textwidth]{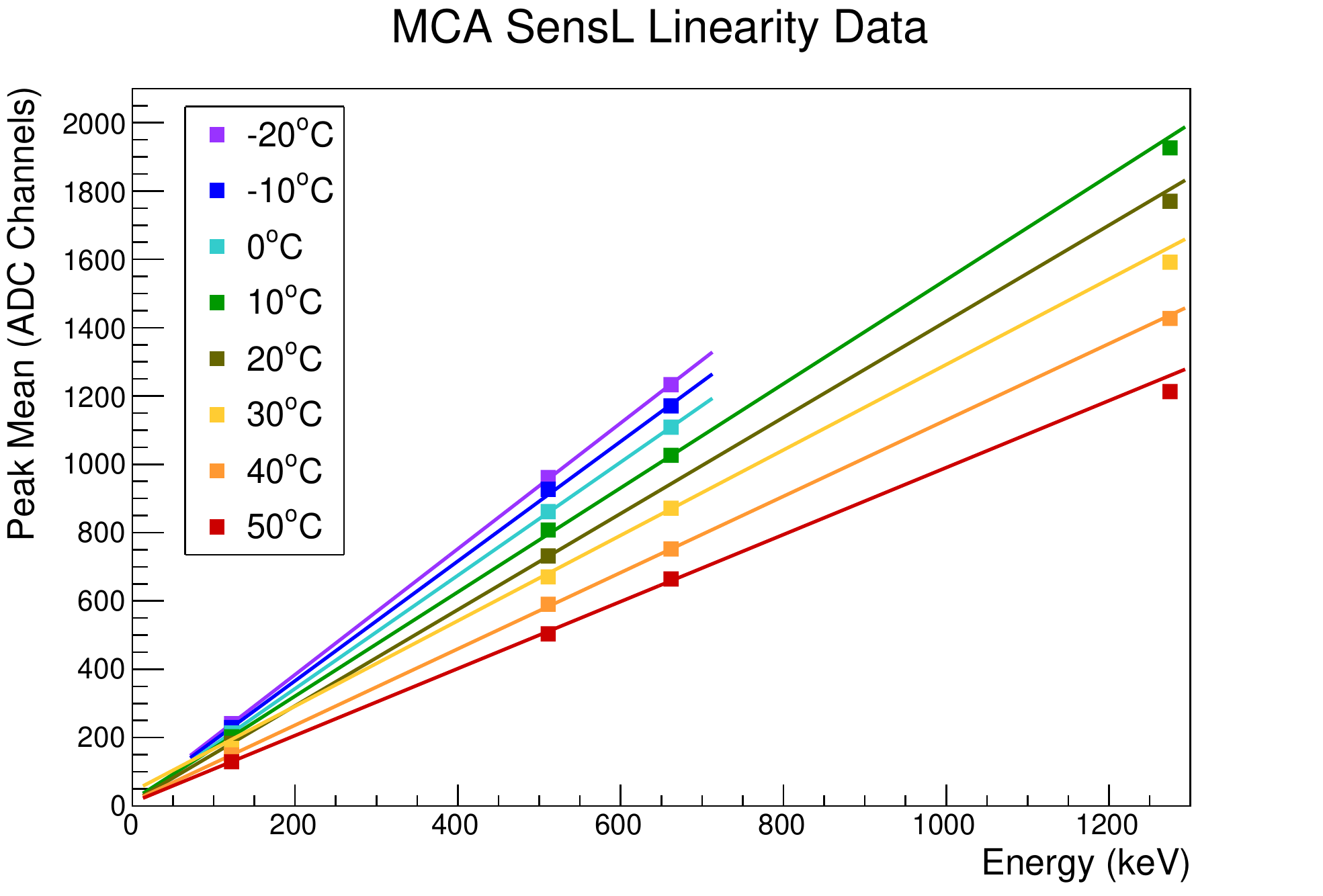}
\caption{Linearity from 122~keV to 1275~keV for the SensL SiPM.}
\label{fig:lins}
\end{figure}

\begin{figure}[h]
\centering
\includegraphics[width=0.42\textwidth]{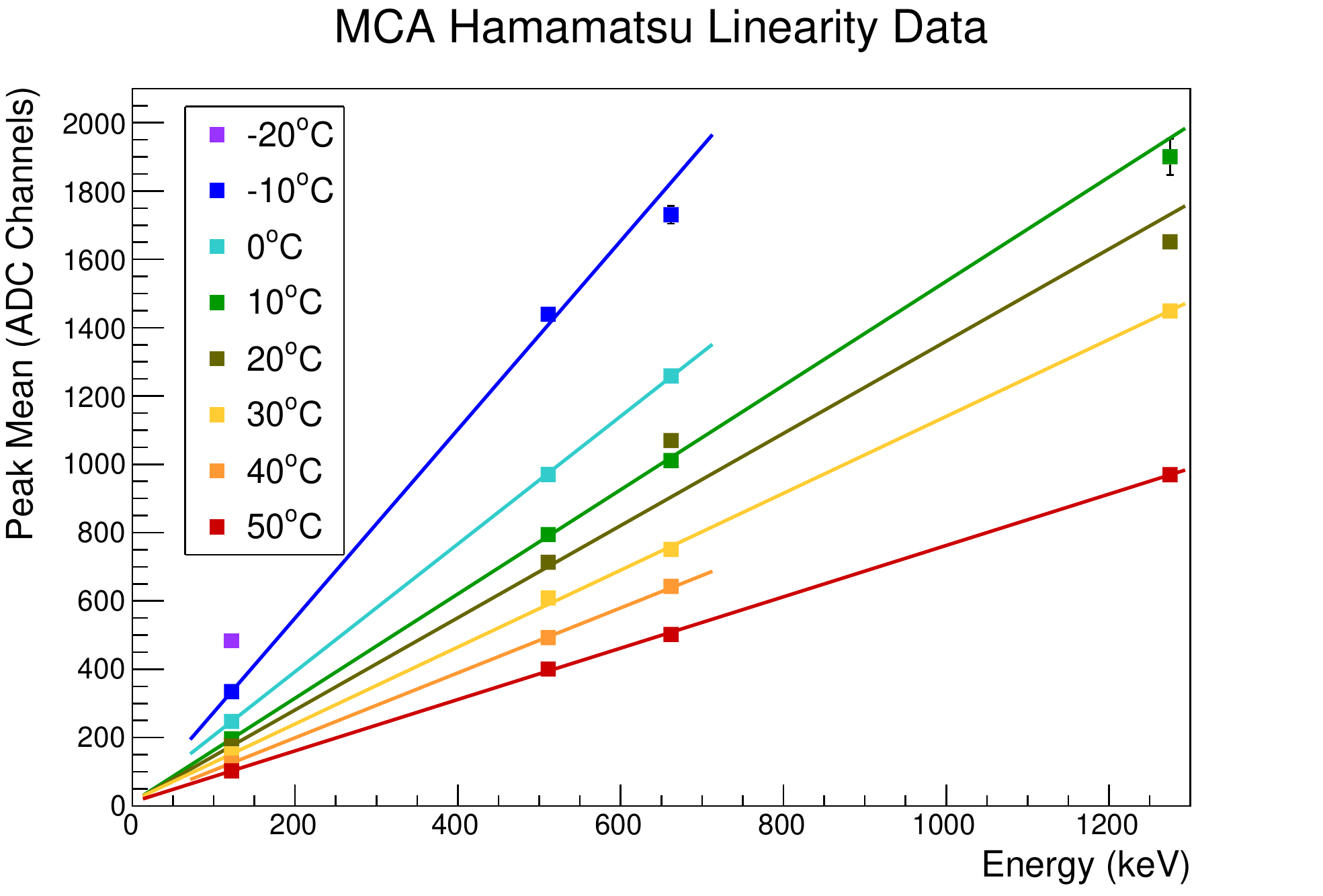}
\caption{Linearity from 122~keV to 1275~keV for the Hamamatsu MPPC.}
\label{fig:linh}
\end{figure}

\begin{figure}[h]
\centering
\includegraphics[width=0.42\textwidth]{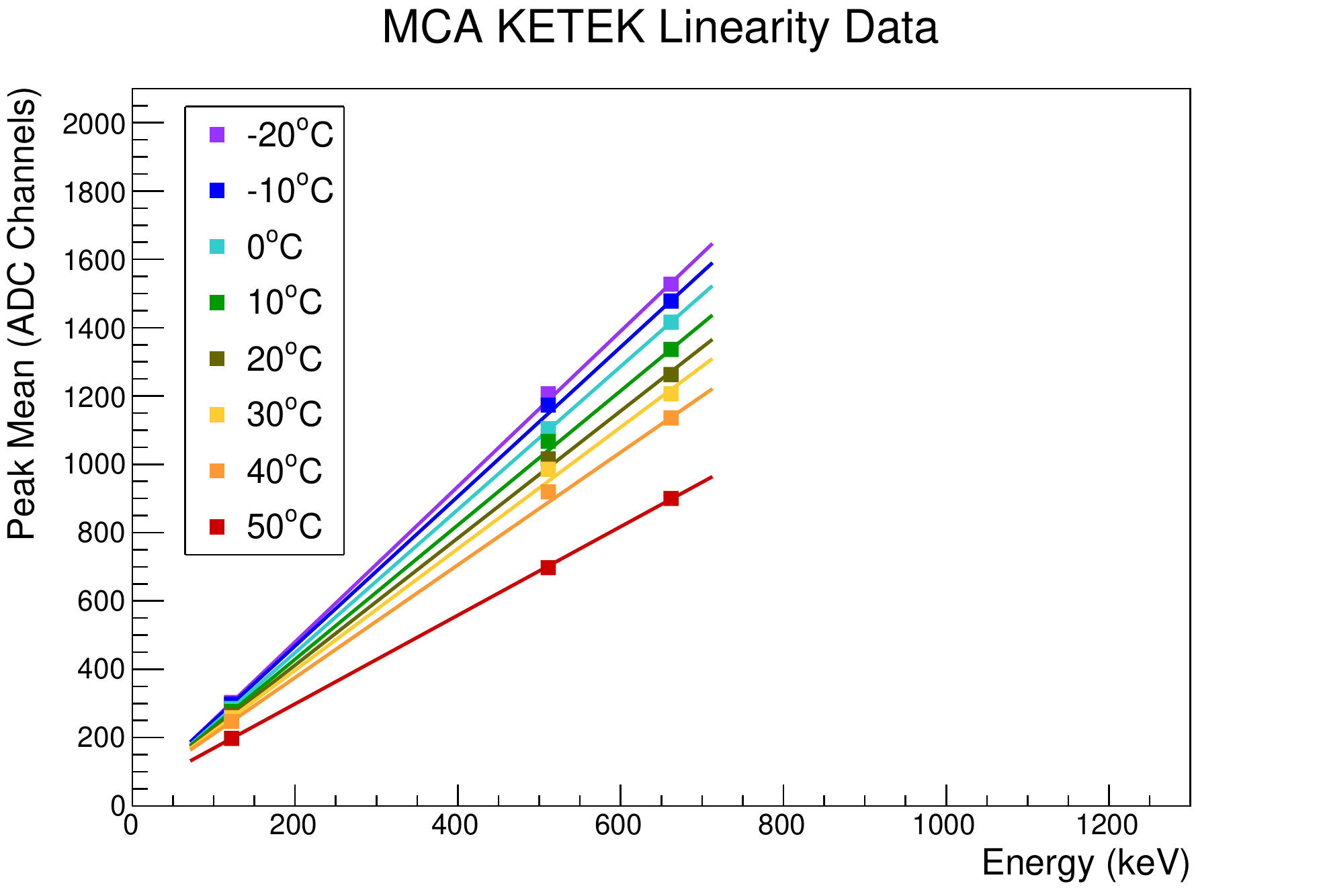}
\caption{Linearity from 122~keV to 1275~keV for the KETEK SiPM.}
\label{fig:link}
\end{figure}

The linearity at each temperature is shown in Figs.~\ref{fig:lins}, \ref{fig:linh}, and \ref{fig:link} for SensL, Hamamatsu, and KETEK, respectively.  In some cases the data point at 1275~keV could not be fit either due to poor statistics or due to being off-scale in the MCA.  All three SSPMs exhibit good linearity, with an $R^2$-correlation coefficient between ADC channel and energy better than 0.997 and in many cases better than 0.999, with the exception of the Hamamatsu MPPC at $-10^{\circ}$C which has an $R^2$ of 0.994 ($R^2$ for $-20^{\circ}$C is not available due to data collection issues).

\begin{figure}[h!]
\centering
\includegraphics[width=0.42\textwidth]{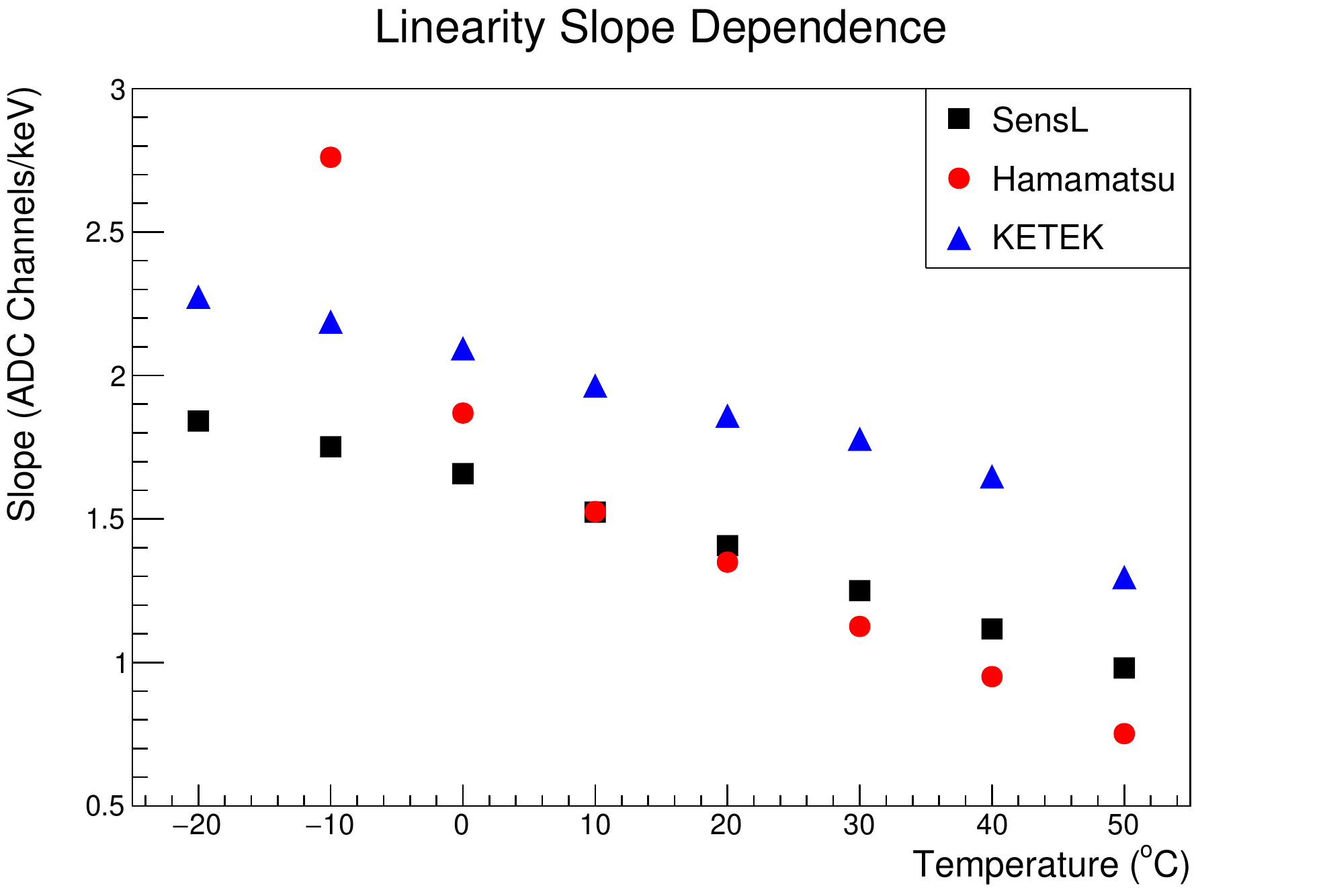}
\caption{Slope in ADC channel versus energy over the temperature range.}
\label{fig:slope}
\end{figure}

Also of importance for applications where temperature variations occur is how much the gain changes over the temperature range.  Figure~\ref{fig:slope} shows the gain (slope of conversion from energy to ADC channel) for each SSPM as a function of temperature.  The SensL SiPM has the least temperature dependence followed by the KETEK SiPM.  KETEK shows a drop in gain at 50$^{\circ}$C, which could be due to temperature limitations of the device.  The Hamamatsu MPPC shows the most temperature dependence, in particular with nonlinear changes in the gain at low temperatures.

\subsection{Energy resolution}

\begin{figure}[h]
\centering
\includegraphics[width=0.4\textwidth]{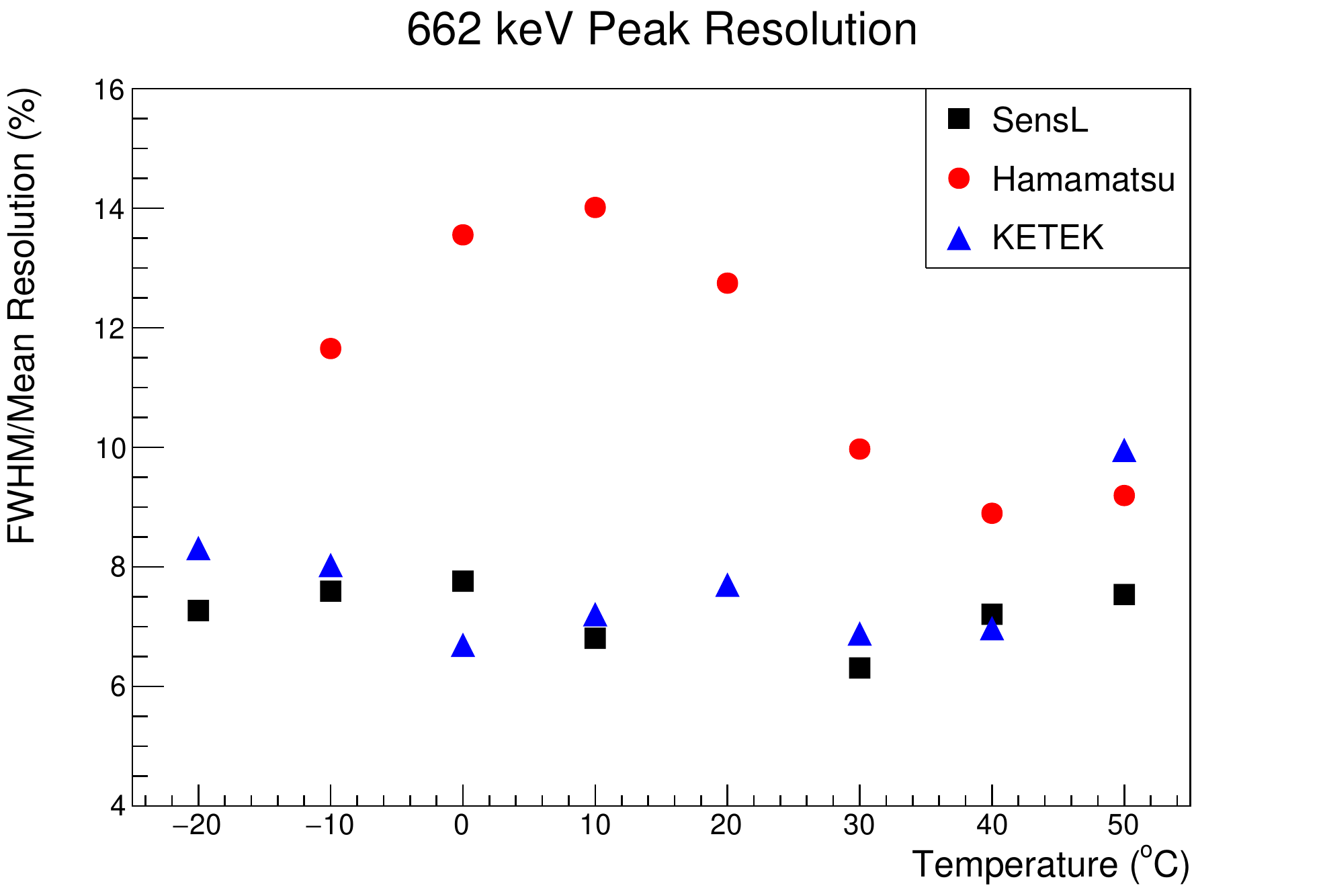}
\caption{FWHM photopeak resolution at 662~keV as a function of SSPM and temperature.}
\label{fig:662}
\end{figure}

The gamma FWHM energy resolution was studied at each of the photopeak energies, but is only shown here for 662~keV in Fig.~\ref{fig:662}.  Room temperature data were also collected, with resolutions at 662~keV of 6.6\%, 8.9\%, and 7.7\% for SensL, Hamamatsu, and KETEK, respectively, similar to or slightly worse than the measured resolutions before the crystals were in-hand.  The temperature cycling data for Hamamatsu suffers from poor statistics and therefore shows significantly worse resolution than the room temperature data, and no conclusions can be made about the temperature dependence for this device.  However, it is clear that for the SensL and KETEK SiPMs there is little dependence of photopeak resolution on temperature at 662~keV.  The exception is the 50$^{\circ}$C data for KETEK, which is likely due to the drop in gain that was previously mentioned.  The resolution for SensL at 20$^{\circ}$C could not be extracted due to an issue with the data.

\subsection{Pulse Shapes}

\begin{figure}[h!]
\centering
\includegraphics[width=0.4\textwidth]{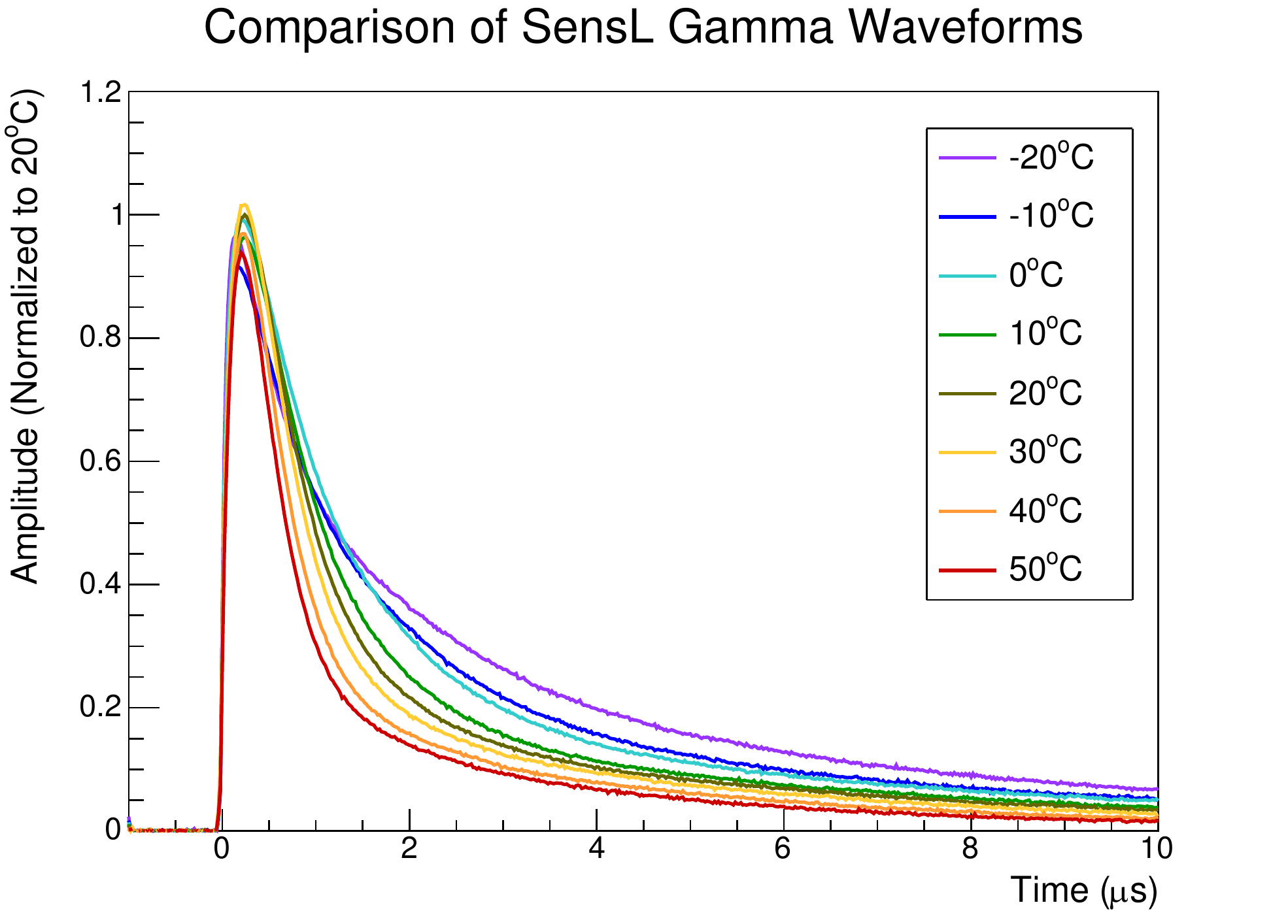}
\includegraphics[width=0.4\textwidth]{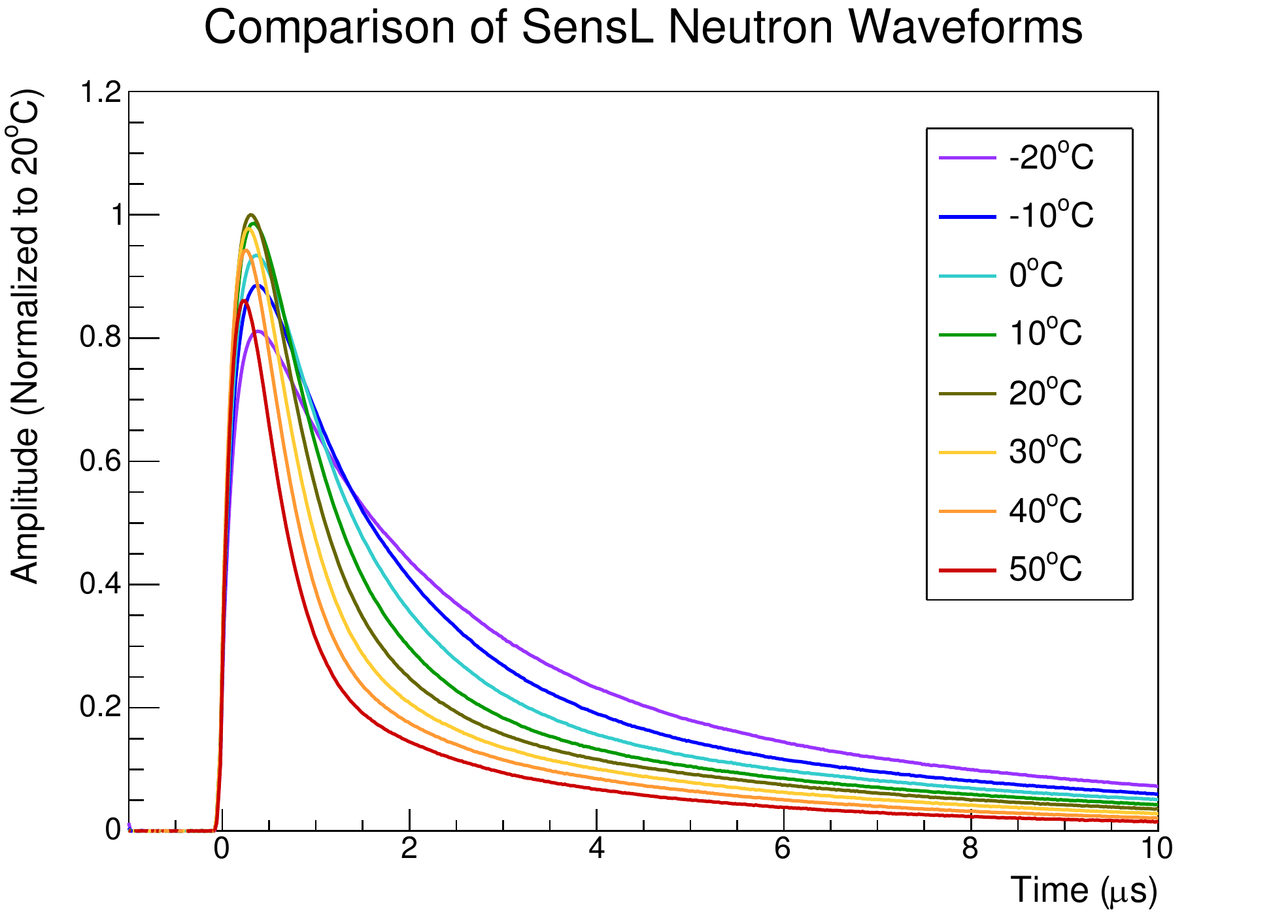}
\caption{Average gamma (top) and neutron (bottom) waveforms for the SensL SiPM as a function of temperature.}
\label{fig:pulsesensl}
\end{figure}

\begin{figure}[h!]
\centering
\includegraphics[width=0.4\textwidth]{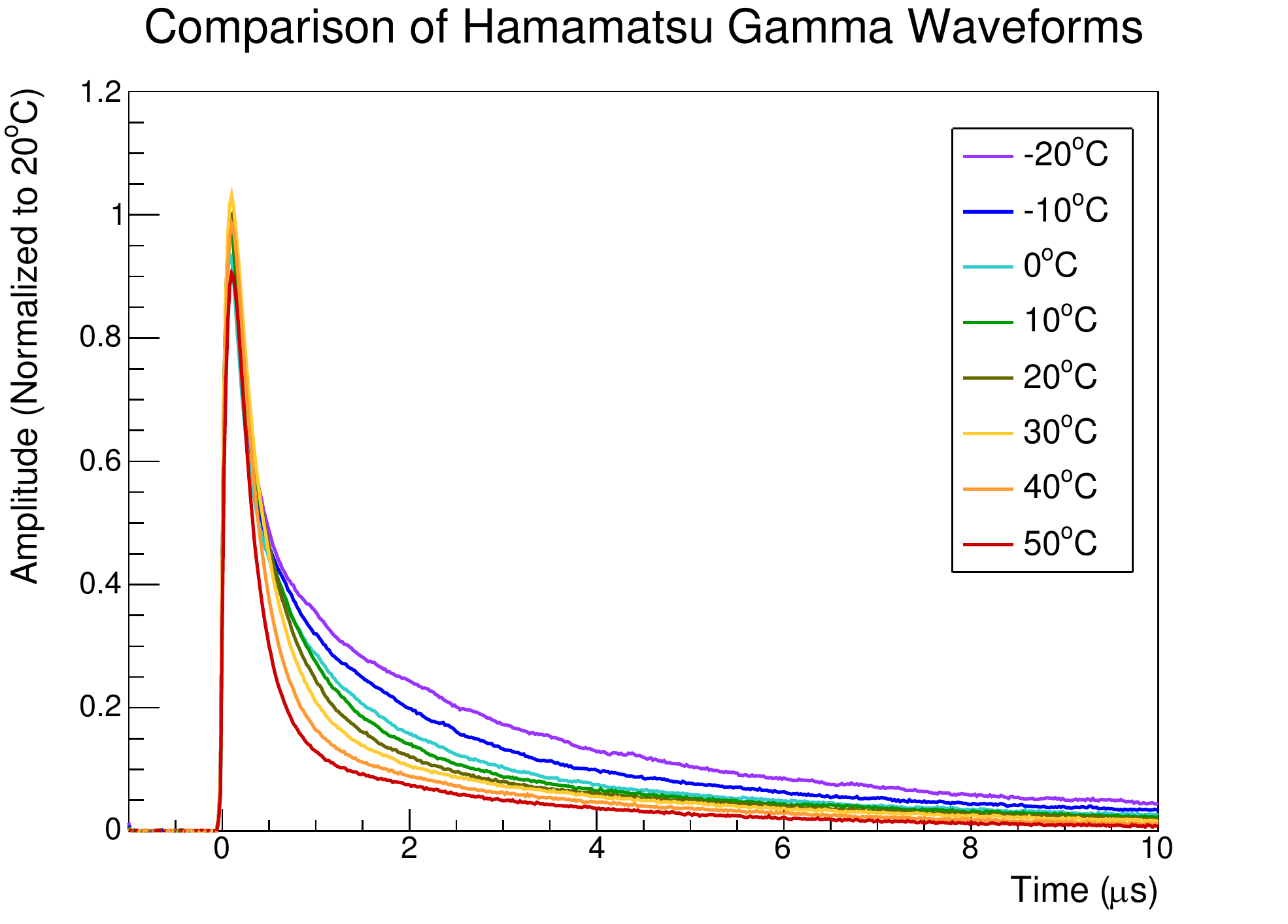}
\includegraphics[width=0.4\textwidth]{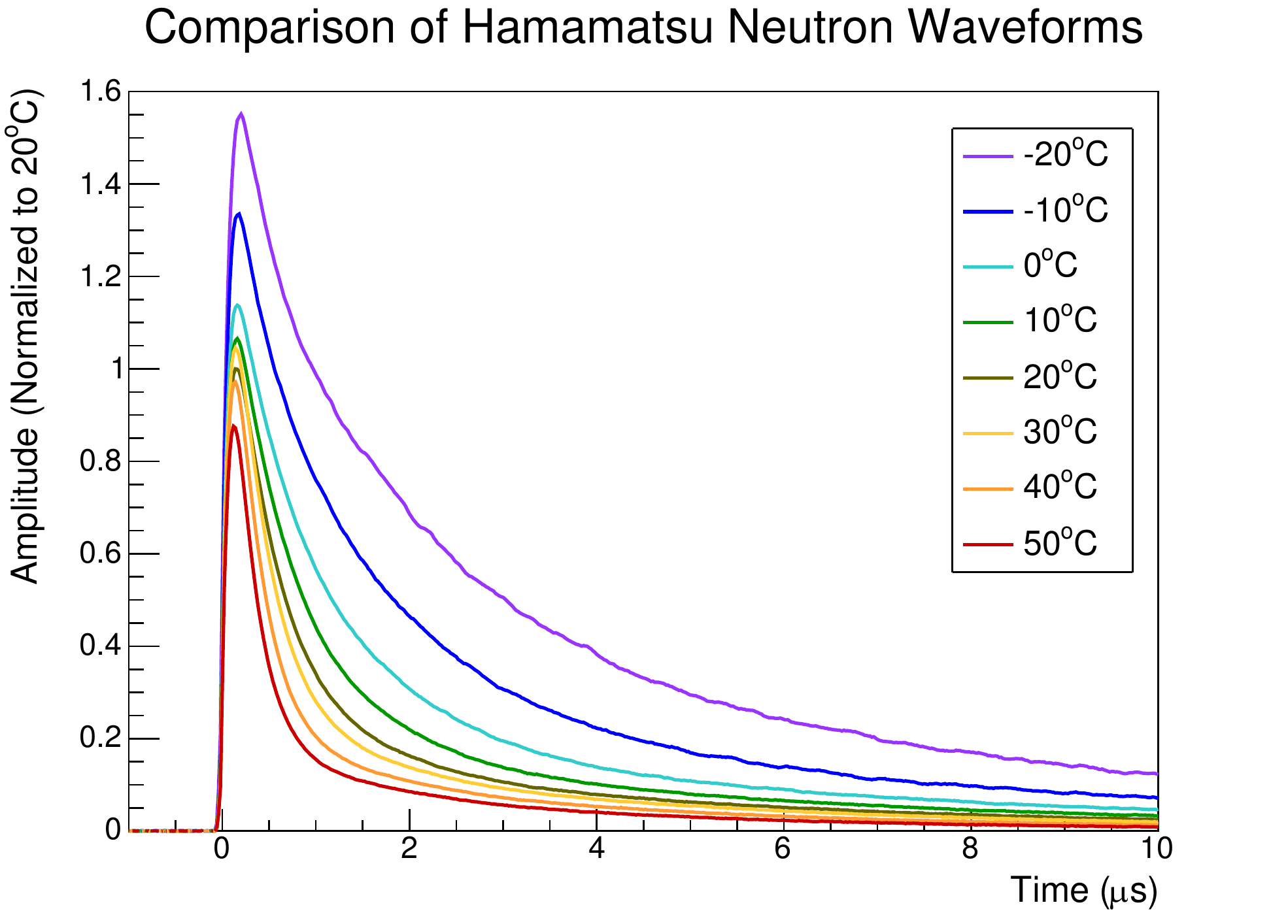}
\caption{Average gamma (top) and neutron (bottom) waveforms for the Hamamatsu MPPC as a function of temperature.}
\label{fig:pulseham}
\end{figure}

\begin{figure}[h!]
\centering
\includegraphics[width=0.4\textwidth]{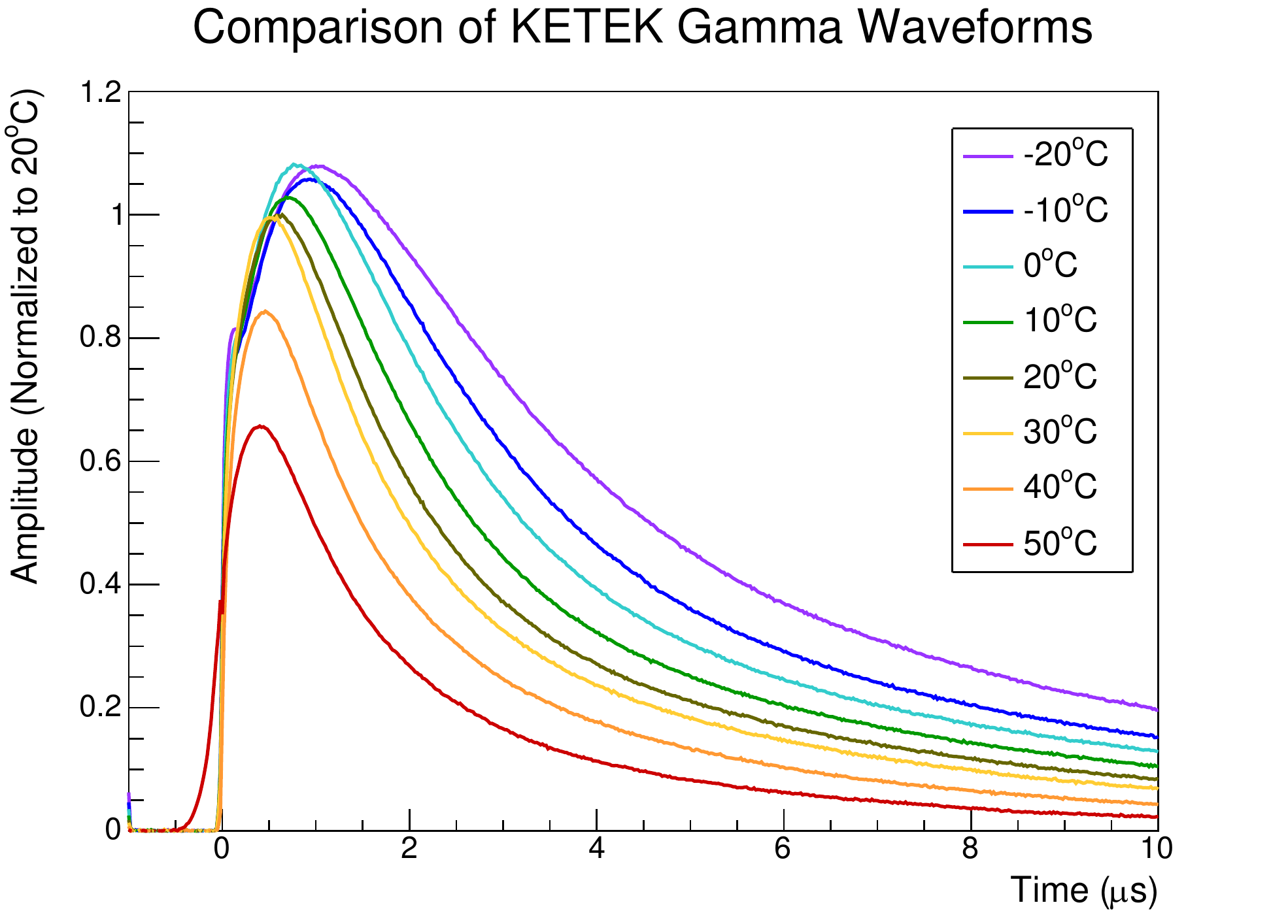}
\includegraphics[width=0.4\textwidth]{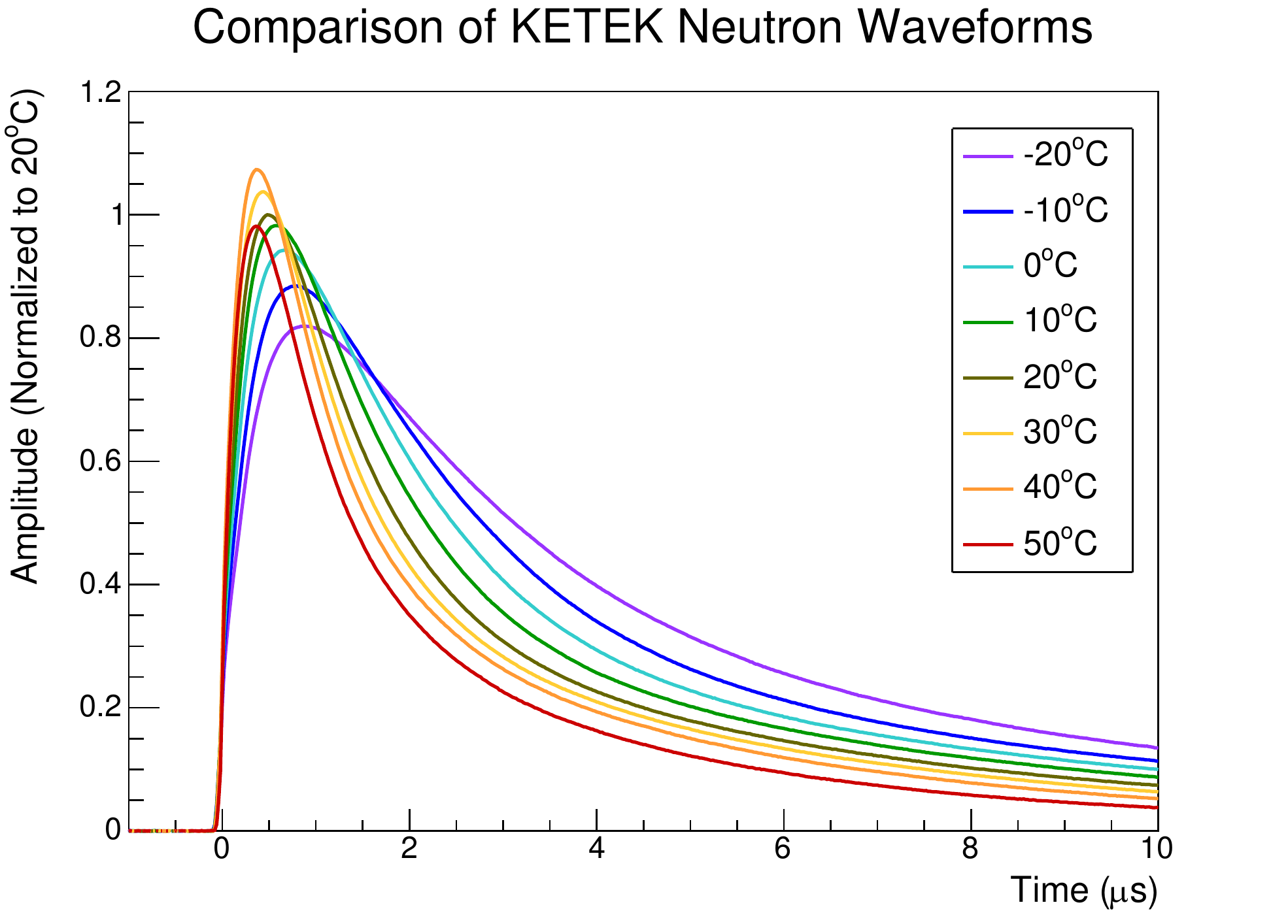}
\caption{Average gamma (top) and neutron (bottom) waveforms for the KETEK SiPM as a function of temperature.}
\label{fig:pulseketek}
\end{figure}

Figures~\ref{fig:pulsesensl}, \ref{fig:pulseham}, and \ref{fig:pulseketek} 
show the average gamma and neutron waveforms for SensL, Hamamatsu, and KETEK, respectively, where each has been normalized to the 20$^{\circ}$C waveform peak.  Note that for this study the same readout electronics were used for each device and no customization was performed at this stage to optimize timing performance.

The gamma waveforms show very little dependence in pulse-height versus temperature, with the exception of the KETEK SiPM which is known to exhibit a drop in gain at the highest temperature as discussed in Section~\ref{sec:lin}.  For both the gamma and neutron waveforms, the time response is slower as the temperature is decreased.  Interestingly, for the KETEK SiPM, which has the slowest response in this setup, hints of a faster component to the gamma response appears at temperatures below 0$^{\circ}$C as evidenced by inflection on the rising edge.  

The neutron pulse-height does have a dependence on temperature, typically with smaller pulse-heights at lower temperatures.  The SensL SiPM does not follow this trend above room temperature, where the pulse-heights start to decrease.  The full-pulse integrated energy for these temperatures still follows a roughly linear trend of smaller energy at higher temperatures, so it could simply be due to a correlation between pulse-height and recovery time.  The KETEK SiPM at 50$^{\circ}$C has a pulse-height similar to 10$^{\circ}$C due to the as-noted drop in gain.  The Hamamatsu MPPC does not follow this general trend whatsoever, instead the pulse-height increases with decreasing temperature.  This is not fully understood at this time, but again could be due to interplay between the pulse-height and recovery time.  The integrated neutron energy over the full 10~$\mu$s window does follow the expected behavior of smaller energy at higher temperature, but exhibits a much more nonlinear trend similar to the observations from the linearity results.

\begin{figure}[h]
\centering
\includegraphics[width=0.4\textwidth]{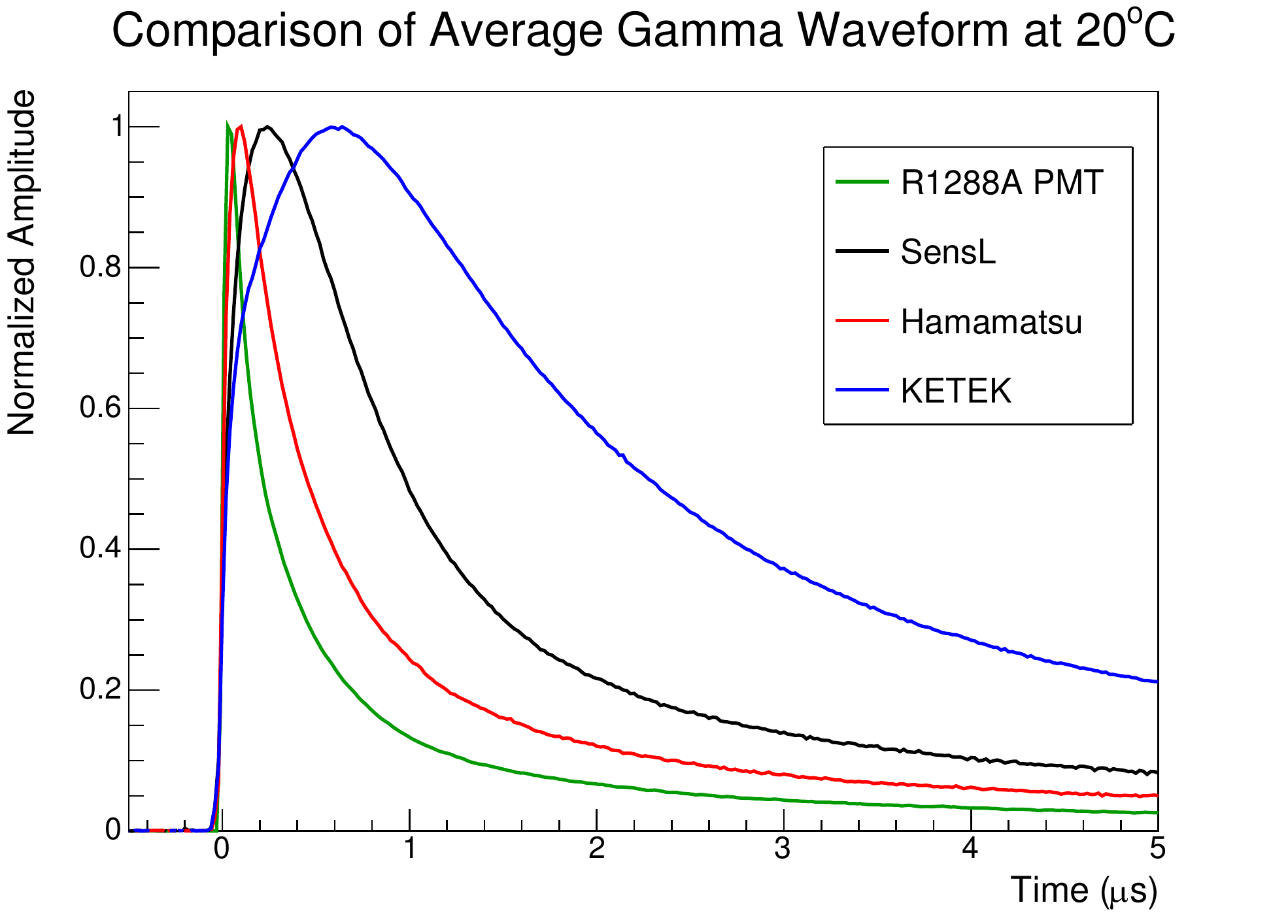}
\includegraphics[width=0.4\textwidth]{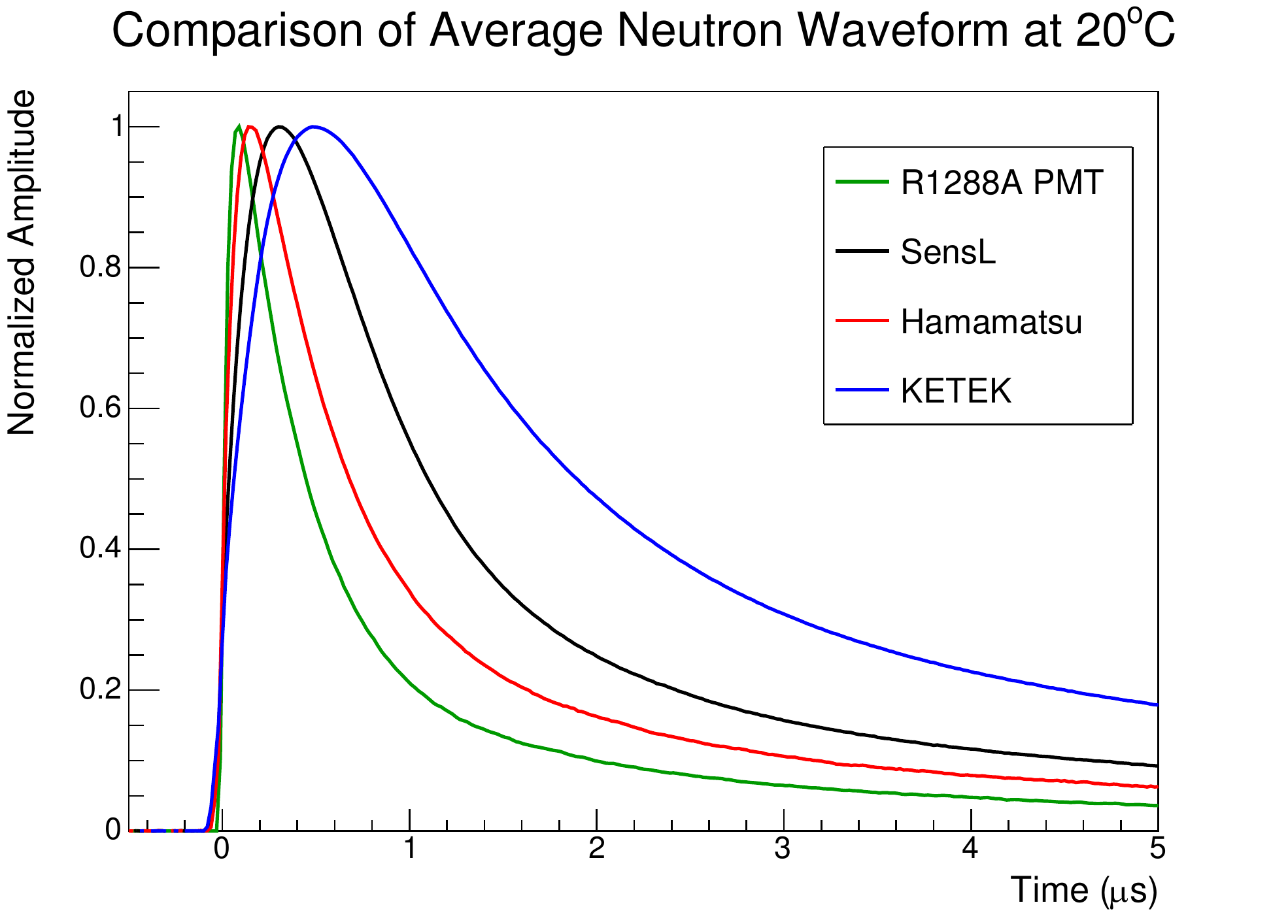}
\caption{Comparison of the average gamma (top) and neutron (bottom) waveforms at 20$^{\circ}$C from the SSPMs and a PMT.}
\label{fig:pmt}
\end{figure}

Finally, a comparison of the pulse shapes can be made by normalizing the pulse heights to unity.  This is shown in Fig.~\ref{fig:pmt} for the three SSPMs at 20$^{\circ}$C with average waveforms from a Hamamatsu R1288A PMT for comparison.  In this simple default comparison with no optimization of readout electronics, the KETEK SiPM showed the slowest time response while the Hamamatsu MPPC was the fastest and most PMT-like.

\subsection{Figure of Merit}

\begin{figure}[h]
\centering
\includegraphics[width=0.4\textwidth]{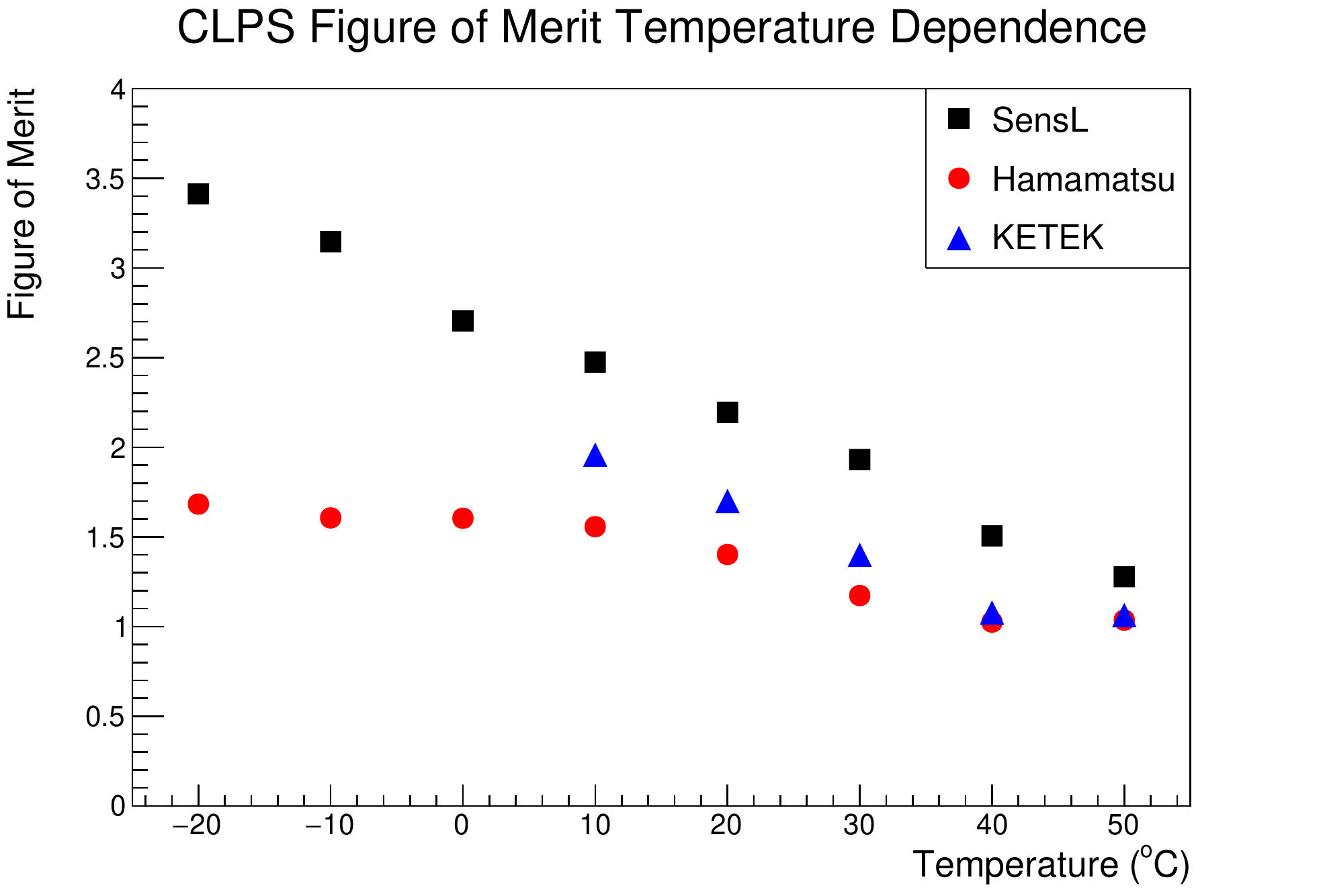}
\caption{Figure of merit for the three SSPMs as a function of temperature.}
\label{fig:fom}
\end{figure}

The figure of merit quantifies the ability to perform PSD and one looks for a FOM being greater than unity.  The FOM versus temperature is shown in Fig.\ref{fig:fom}.  PSD was not possible for the KETEK SiPM below 10$^{\circ}$C due to a data acquisition issue of the neutron peak being off scale.  All three SSPMs exhibit improved FOM at lower temperatures.  The FOM was always better than 1, and was the best for SensL.  At 20$^{\circ}$C the FOM for SensL, Hamamatsu, and KETEK was 2.2, 1.4, and 1.7, respectively.

\section{Summary}

The temperature dependence of three SSPMs coupled to 1~cm$^3$ CLYC scintillators was studied.  The SensL and KETEK SiPMs exhibited excellent linearity over the temperature range of $-20^{\circ}$C to $+50^{\circ}$C.  The Hamamatsu MPPC showed a nonlinearity in gain at the lower temperatures.  Statistical limitations of the dataset affected our ability to extract meaningful resolution results, however, it does appear that the SensL and KETEK SiPM energy resolution at 662~keV does not depend on temperature.  Finally, the SSPM and CLYC response was generally slower at lower temperatures for all three devices, but PSD was performed successfully for each device over the full temperature range studied, with generally better FOM at lower temperatures and the best FOM overall for the SensL SiPM.

%
%

\end{document}